\documentclass[aps,prb,twocolumn,superscriptaddress]{revtex4-2}

\usepackage{amsmath}
\usepackage{graphicx}
\usepackage{hyperref}
\usepackage{siunitx}
\usepackage{todonotes}
\usepackage{xspace}
\usepackage{amsfonts}
\usepackage{bm}
\usepackage{placeins}

\newcommand{\FF}{F\!F}
\newcommand{\jsc}{j_\mathrm{sc}}
\newcommand{\Voc}{V_\mathrm{oc}}

\begin{document}

\title{Visualizing the Link Between Nanomorphology and Energetic Disorder in 3D Organic Solar Cells}

\author{Pelin \c{C}ilo\u{g}lu}
\affiliation{Faculty of Mathematics, Technische Universität Chemnitz, Reichenhainer Straße 39, D-09107 Chemnitz, Germany}

\author{Carmen Tretmans}
\affiliation{Institute of Mathematics, University of Augsburg, Universitätsstraße 12a, 86159 Augsburg, Germany}

\author{Carsten Deibel}
\affiliation{Institut für Physik, Technische Universität Chemnitz, Chemnitz 09126, Germany}

\author{Jan-F. Pietschmann}
\affiliation{Institute of Mathematics and Centre for Advanced Analytics and Predictive Sciences (CAAPS), University of Augsburg, Universitätsstraße 12a, 86159 Augsburg, Germany}

\author{Martin Stoll}
\affiliation{Faculty of Mathematics, Technische Universität Chemnitz, Reichenhainer Straße 39, D-09107 Chemnitz, Germany}

\author{Roderick C. I. MacKenzie}
\affiliation{Department of Engineering, Durham University, Durham, United Kingdom}

\date{\today}

\begin{abstract}
The performance of organic bulk heterojunction (BHJ) solar cells is highly sensitive to both nanomorphology and energetic disorder arising from microscopic molecular packing and structural defects. However, most models used to understand these devices are either one-dimensional effective medium approximations that neglect spatial and energetic disorder or three-dimensional Monte Carlo simulations that are computationally intensive.

In this work, we present the results from a three-dimensional hybrid model capable of operating at both high carrier densities and incorporating the effects of energetic disorder. We first generate realistic morphologies using a phase-field approach that accounts for solvent evaporation during film formation. Using these example morphologies, we systematically study the interplay between energetic disorder and configurational disorder at carrier densities representative of real device operation. This enables us to separate and visualize the impact of the nanomorphology and energetic disorder on device performance.

Our results reveal that, even when macroscopic percolation pathways remain intact, energetic disorder limits performance primarily through suppressed charge extraction in interconnected domains. This suggest that optimizing molecular packing at the nanoscale is as critical as controlling phase separation at the mesoscale, highlighting the need for multiscale design strategies in next-generation BHJ devices.

\end{abstract}

\maketitle

\section{Introduction}

\begin{figure}
    \centering
    \includegraphics[width=\columnwidth]{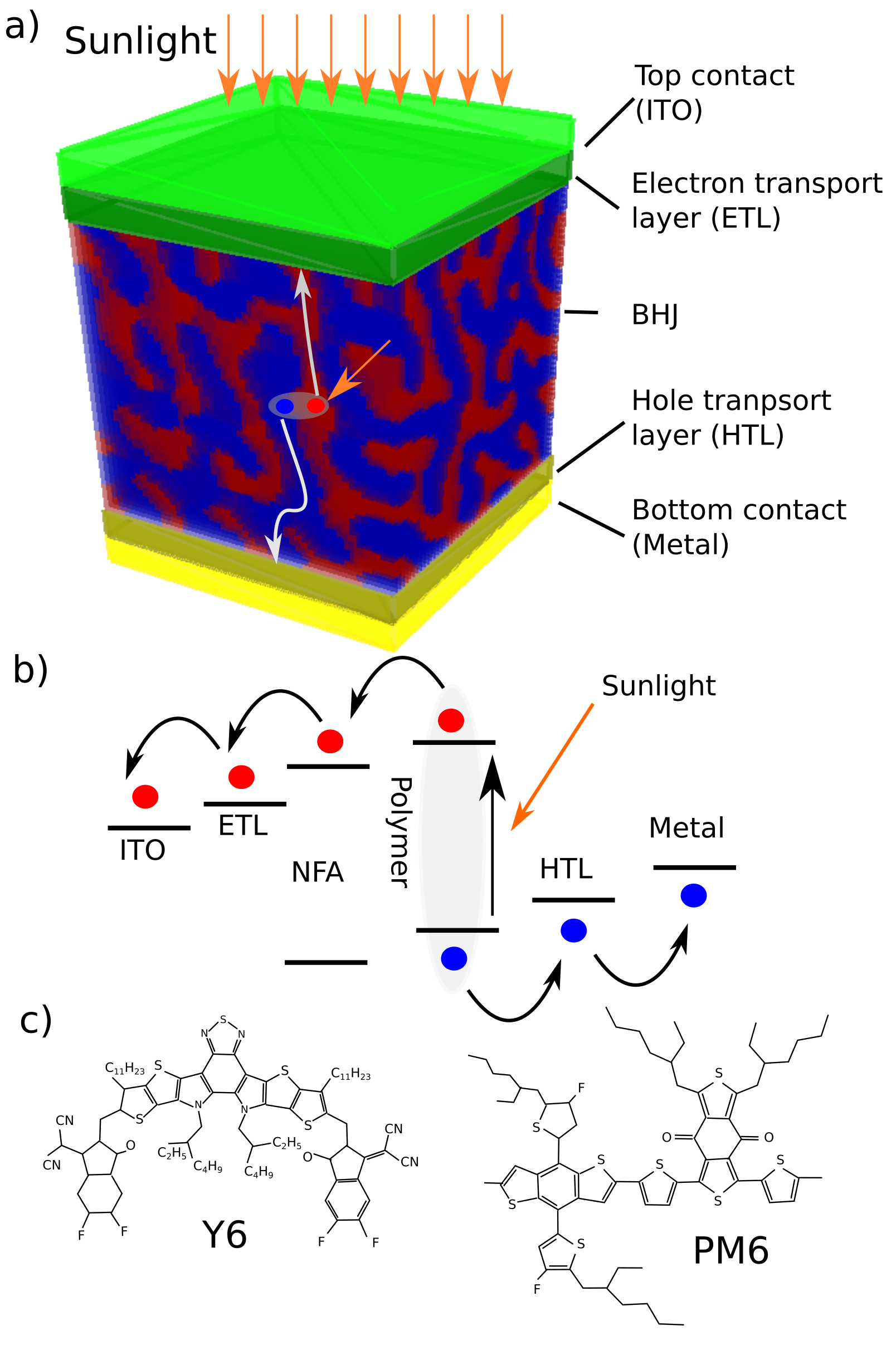}
    \caption{a) The structure of a typical BHJ solar cell, a mixture of red (acceptor) and blue (donor) materials: light absorption creates excitons, which diffuse to the donor--acceptor interface. By charge transfer, electrons and holes are generated from the excitons, which then move by drift and diffusion to the corresponding electrodes; b) The band structure of a typical BHJ solar cell; c) In this paper our device architecture and performance closely follow that of the PM6:Y6 cell. The corresponding chemical structures are shown.}
    \label{fig:device_figure}
\end{figure}

Organic semiconductors hold significant promise for next-generation photovoltaic devices, offering lightweight, flexible, and potentially low-cost alternatives to their inorganic counterparts. Recent polymer--non-fullerene acceptor devices have demonstrated power conversion efficiencies (PCE) of around $\SI{20}{\percent}$ \cite{Zhu2022,Zhu2024,Dong2025}. Most high-performance devices rely on bulk heterojunction (BHJ) architectures \cite{https://doi.org/10.1002/adfm.202414941}, which consist of a nanoscale blend of electron donor and acceptor materials sandwiched between two contacts. This structure is visible in Figure~\ref{fig:device_figure}a. The device works as a solar cell by the donor or the acceptor absorbing light and generating strongly bound excitons, which diffuse to the donor--acceptor interface where the energetic offset at the interface dissociates them by charge transfer. At this point the charge carriers move by drift and diffusion to the corresponding electrodes: the hole through the donor material and the electron through the acceptor material. The energetics of this process are shown in Figure~\ref{fig:device_figure}b. The performance of BHJ devices is very sensitive to the detailed nanomorphology of the donor--acceptor blend \cite{kniepert_conclusive_2014}. A BHJ with well-defined polymer--acceptor transport pathways to the contacts will generally provide a higher cell efficiency than a device with isolated islands of either polymer-in-acceptor or vice versa, which would promote charge trapping and subsequent recombination.

Although the spatial configuration of the donor--acceptor phase separation of the BHJ on the 10--50~nm length scale is important, microscopic packing of the molecules and how well-coupled frontier orbitals is also very important.\cite{Zhang2024} Non-perfect packing on the molecular level as well as the introduction of impurities leads to the formation of an energetic distribution of trap states below the transport levels. These traps can hold a significant amount of charge, and can act as recombination centers that reduce; the short circuit current density $\jsc$; the open circuit voltage $\Voc$;  the fill factor $\FF$; and the PCE. Understanding and controlling the nanomorphology is therefore critical to optimizing charge transport, reducing recombination losses, and achieving high-efficiency organic photovoltaic devices.

Traditional approaches to modeling charge transport in these systems often rely on one-dimensional effective medium approximations, that reduce the complexities of the BHJ to a classical 1D inorganic diode model while neglecting both the complex nanomorphology and microscopic energetic disorder \cite{PhysRevB.94.035205}.   While these models offer valuable insights into fundamental transport mechanisms; are computationally very efficient; and can be extended to include energetic disorder \cite{Mackenzie2012extractingmicroscopic}, they struggle to capture the full three-dimensional connectivity and percolation effects \cite{https://doi.org/10.1002/pssb.2221750102} that govern charge transport in real devices. Some notable 2D efforts have been made to include spatial separation of donor and acceptor \cite{Buxton2006,Stelzl2012}, however these models did not include energetic disorder.

Another approach is to perform full Monte Carlo (KMC) simulations. KMC can be a powerful tool to study disorder and morphological configuration \cite{Heiber2016,upreti2021slow,wilken2020}, however the simulations are slow due to the need to consider the hopping of individual electrons in the energetically and spatially disordered landscape. As the carrier density rises to realistic levels, the computational cost of the simulations increases due the cost of computing carrier--carrier interactions. This means that modeling devices at 1~Sun (AM1.5G spectrum at 100~mW/cm$^2$) with realistic injecting contacts is difficult. These limitations have hampered the use of KMC models in the field.

In this work, we address this gap between full KMC and drift--diffusion simulations by developing a hybrid modeling framework that incorporates both energetic disorder and a realistic three-dimensional nanomorphology. We begin by generating BHJ morphologies using a phase-field model that captures the formation of donor--acceptor phase separation using a three-component model of donor, acceptor, and solvent. Through virtual evaporation, we are able to arrive at a final morphology. These morphologies are then used as input into our hybrid model.

This approach allows us to decouple the influence of configurational and energetic disorder and identify how each component of disorder affects key device parameters such as short-circuit current ($\jsc$ ), open-circuit voltage ($\Voc$), and fill factor ($FF$). Our results reveal that energetic disorder limits performance primarily through suppressed charge extraction in interconnected domains, even when macroscopic percolation pathways remain intact. This highlights that optimizing molecular packing at the nanoscale is as critical as controlling phase separation at the mesoscale.

\section{The model}

\subsection{Morphology generation and device structure}

The device studied is broadly based on the PM6:Y6 material system and has the structure Glass/ITO/SnO$_2$/PM6:Y6/MoO$_3$/Ag\cite{Woepke2022transport}. This is visible in Figure~\ref{fig:device_figure}a. PM6:Y6 was chosen because, with the emergence of non-fullerene acceptors, it has become a key model system in the field \cite{Shoaee2023}. Our device was given a box shaped active layer of size 110~nm $\times$ 110~nm $\times$ 110~nm. This size was chosen because a film thickness of 70--150~nm is commonly used in experiments and -- with a strong built-in field normal to the contacts -- transport is limited horizontally, reducing the need for a much wider simulation window.

\begin{figure}
    \centering
    \includegraphics[width=\columnwidth]{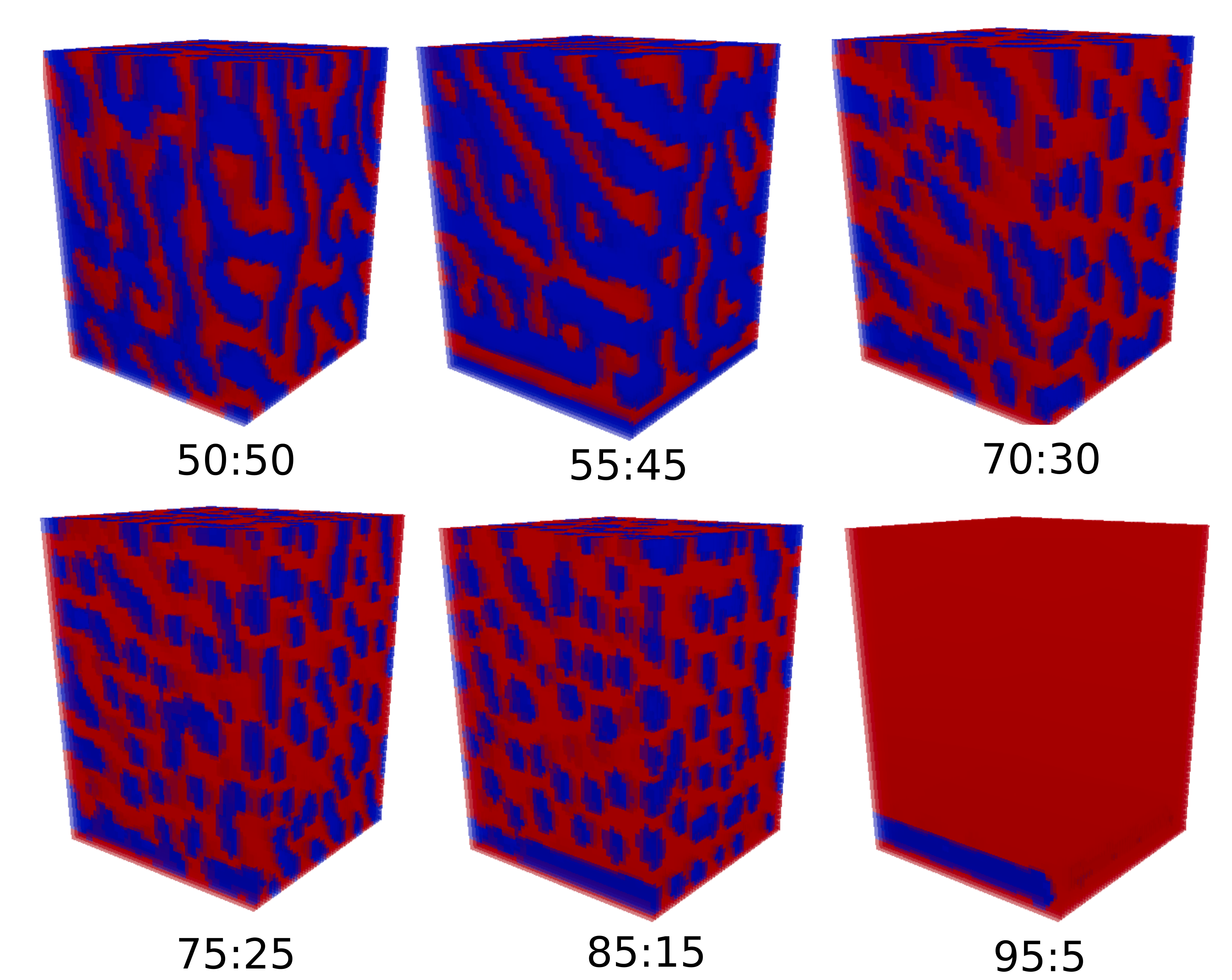}
    \caption{Generated morpholgies with varying acceptor:donor ratios.}
    \label{fig:morphology_figure}
\end{figure}

To model the formation of the BHJ, we simulate a three-component system consisting of a donor, an acceptor, and a solvent. During the drying process, the solvent evaporates, leading to phase separation of the donor and acceptor materials into distinct domains. To capture this complex interplay of processes, we use a Cahn--Hilliard-type phase-field model \cite{KBergermann_2023, PCiloglu_2025} to simulate the temporal evolution of the domains representing the donor and acceptor materials within the blend. The morphology formation is governed by the following Cahn--Hilliard equation:
\begin{equation}
    \frac{\partial \phi_i}{\partial t} = \nabla \cdot \left( \gamma_i \nabla \mu_i \right), \quad 
\mu_i = \frac{\partial f(\boldsymbol{\phi}) }{\partial \phi_i} - \beta_i \nabla^2 \phi_i,
\label{eq:CH}
\end{equation}
where $\phi_i$, $i \in \{ p, nfa, s \}$, denote the volume fraction of polymer, non-fullerene acceptor (NFA), and solvent. Here, $\mu_i$ is the chemical potential and $f(\boldsymbol{\phi})$ is the local part of the free energy density, where $\boldsymbol{\phi}$ denotes the set of all three volume fractions. The constant parameters $\gamma_i$ and $\beta_i$ represent the molecule mobility and an interface parameter related to the width of the transition layers, respectively. Six example morphologies can be seen in Figure~\ref{fig:morphology_figure}. More details on the morphology generation can be found in the SI, section~\ref{ssec:SI_morphology}.

\subsection{Charge transport}\label{sec:ElecMod}

\begin{figure*}
    \centering
    \includegraphics[width=\textwidth]{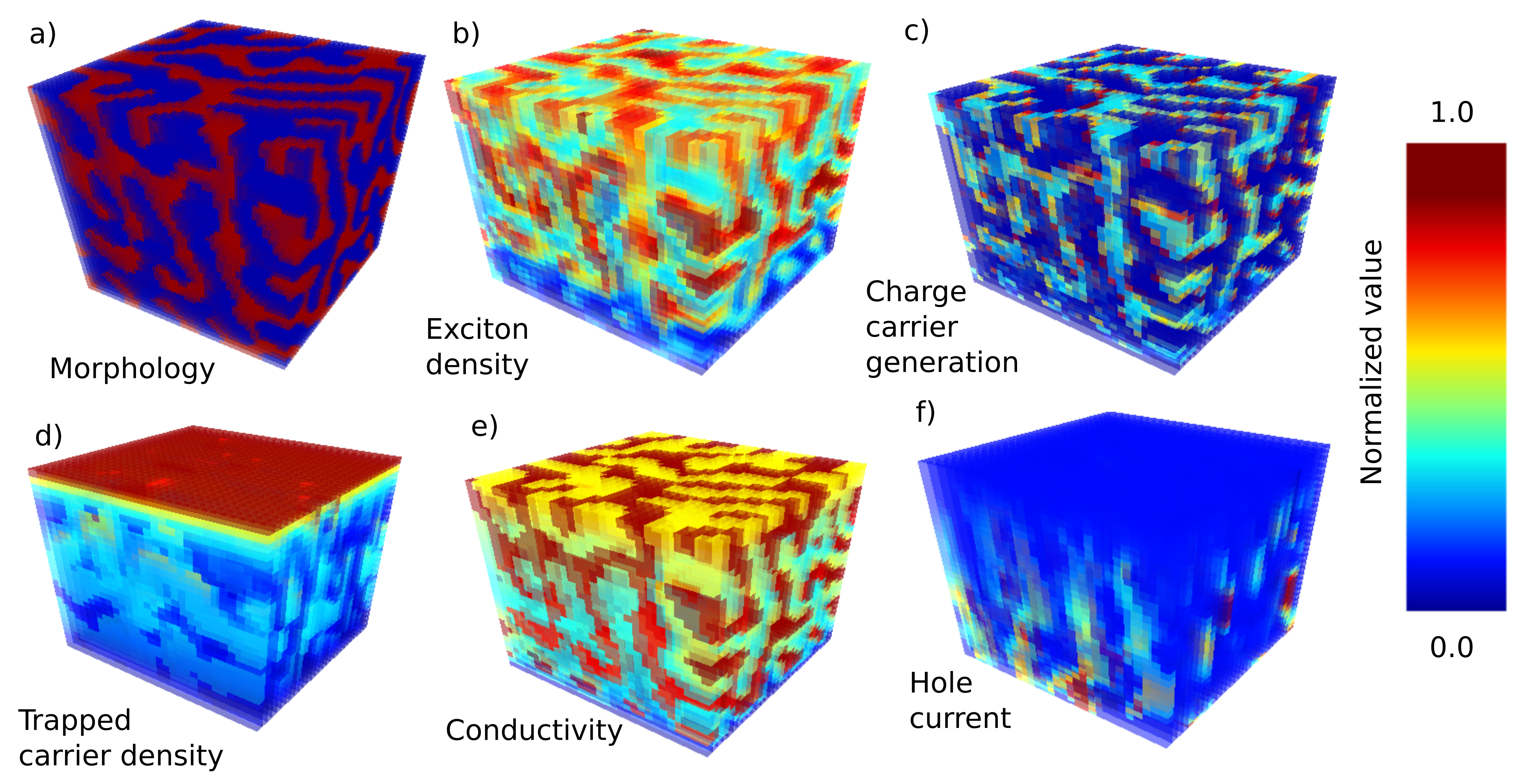}
    \caption{a) The morphology generated by the phase-field model; b) The exciton density generated by incident light; c) Charge carrier generation at the edge of the domains; d) Trapped hole carrier density; e) Conductivity; f) Hole current flowing normal to the contacts.}
    \label{fig:figure1}
\end{figure*}

To model charge transport in these complex systems and to understand the effects of configurational disorder, the drift--diffusion equations are solved in quasi-3D (see later and SI for full details). Figure~\ref{fig:figure1} shows a single simulation run of the electrical model, with Figure~\ref{fig:figure1}a showing the input morphology. The AM1.5G solar spectrum is simulated propagating through the ITO substrate normal to the interface using a transfer matrix approach. Each component of the blend along with the electrode layers is assigned an experimental absorption profile as a function of wavelength, enabling absorption as a function of wavelength to be modeled. Light scattering away from the normal is not considered. The absorbed photons generate excitons (Figure~\ref{fig:figure1}b) which are allowed to diffuse/recombine/split according to the equation:
\begin{equation}
    \label{eq:exciton}
    \frac{\partial X}{\partial t} = \nabla \cdot D \nabla X +G_{optical} -k_{dis} X -k_{FRET} X- k_{PL} X-\alpha X^2.
\end{equation}
Here, $X$ is the exciton density, $D$ the diffusion coefficient, and $G_{optical}$ the exciton generation rate. The exciton dissociation rate to free charge carriers is given by $k_{dis}$, $k_{FRET}$ is the F\"{o}rster resonance energy transfer rate constant, $k_{PL}$ the prefactor describing the radiative loss of excitons, and $\alpha$ an exciton--exciton annihilation rate constant.

The value of $k_{dis}$ is set to zero throughout the device, except within a 10~nm region adjacent to the donor--acceptor interfaces. This approach allows excitons to be generated everywhere in the device, but ensures that only those near an interface can dissociate and contribute to the photocurrent. As a result, excitons created in large domains far from interfaces have a lower probability of contributing to the photocurrent and otherwise recombine geminately. This yields a three-dimensional map of the charge carrier generation rate, $k_{dis}$, concentrated near the 3D interfaces (Figure~\ref{fig:figure1}c)

Once the photogenerated charge distribution has been generated, it is fed into a 3D drift--diffusion model with an exponential tail of trap states. The model is based on our previously developed simulation framework (OghmaNano) \cite{Mackenzie2020ohmicspacecharge, Mackenzie2012extractingmicroscopic, oghmanano}.
%Before ending our discussion on generation, it should be noted that although the textbook description of an organic solar cell always described light being absorbed and excitons generated in the donor, in modern materials as much (if not more) light can be absorbed in the acceptor -- this is accounted for by using measured n/k data.

The model solves Poisson’s equation to account for electrostatic effects,
\begin{equation}
    \nabla \cdot [\epsilon(\mathbf{r}) \nabla \phi(\mathbf{r})] = -q [p_{f/t}(\mathbf{r}) - n_{f/t}(\mathbf{r})],
\end{equation}
where $\phi$ is the electrostatic potential, $\epsilon(\mathbf{r})$ is the position-dependent dielectric constant, $q$ is the elementary charge, $n_{f/t}(\mathbf{r})$ is the sum of the free electron density and the trapped electron density, and similarly $p_{f/t}(\mathbf{r})$ is the hole density. While charge transport is described by the electron and hole continuity equations:
\begin{align}
    \frac{\partial n_f}{\partial t} &= \nabla \cdot [\mu_n n_f \nabla \phi + D_n \nabla n_f] - R(n_{f/t},p_{f/t}) + X \cdot k_{dis}, \\
    \frac{\partial p_f}{\partial t} &= -\nabla \cdot [\mu_p p \nabla \phi - D_p \nabla p_f] - R(n_{f/t},p_{f/t}) + X \cdot k_{dis},
\end{align}
where $\mu_{n,p}$ are the electron and hole mobilities, $D_{n,p}$ are the corresponding diffusion coefficients, $R(n_{f/t},p_{f/t})$ represents the loss of free carriers due to free-to-free recombination and free-to-trap recombination, and also carrier trapping. Example current flow can be seen in Figure~\ref{fig:figure1}f.

\subsection{Modeling trapped charge}
\begin{figure}
    \centering
    \includegraphics[width=0.5\columnwidth]{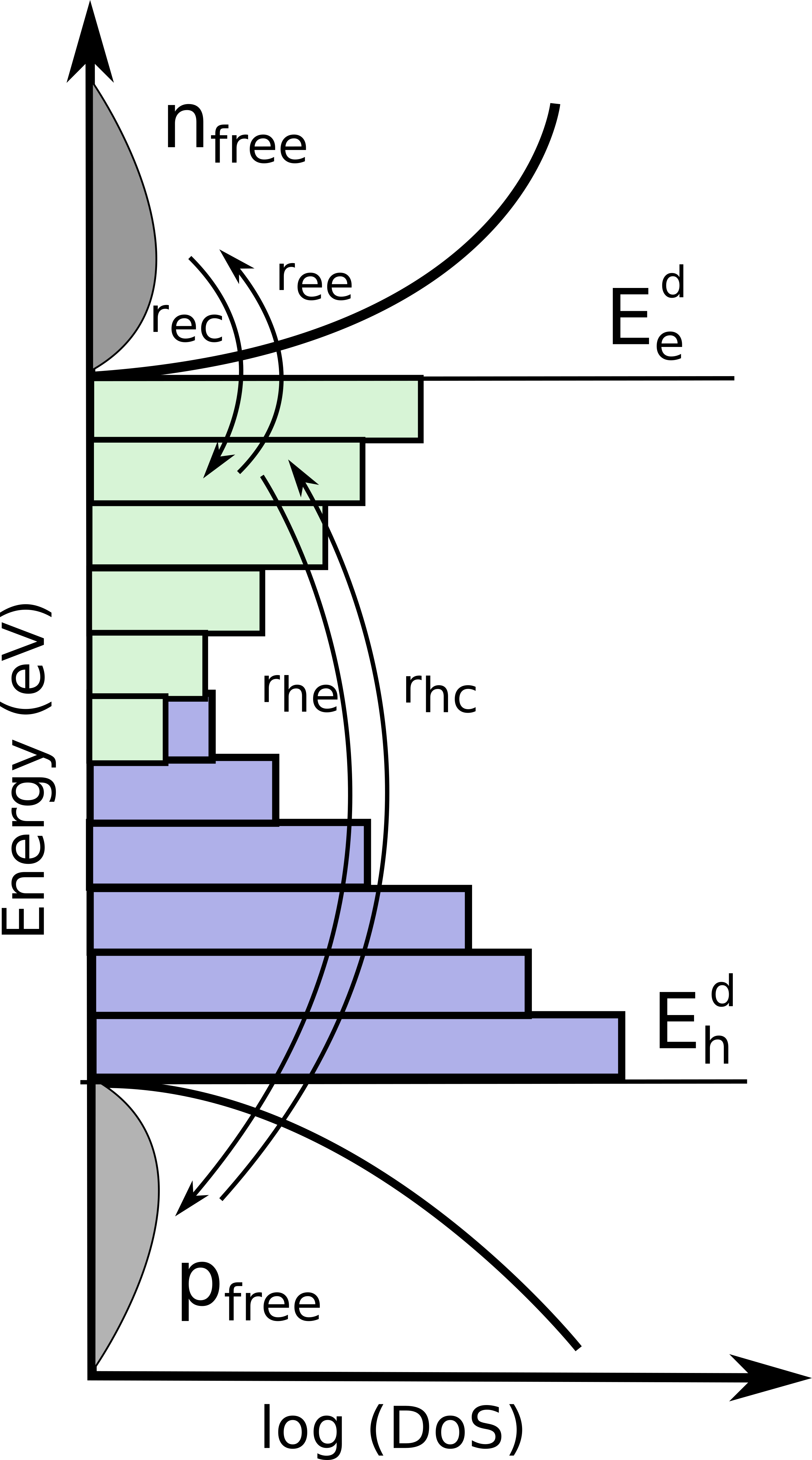}
    \caption{A diagram of the band structure of the model, it consists of free carriers under which trap states reside. Both electrons and holes can become trapped into trap states as described by the Shockley--Read--Hall formulism.}
    \label{fig:figureX}
\end{figure}

Modeling charge carrier trapping is essential for accurately describing transport in disordered semiconductors such as those based on polymers and  small molecules. Standard drift--diffusion models without trap states fail to reproduce the correct dependence of mobility and recombination rate on voltage and carrier density. In our model we break up the distribution of trapped states into a series of independent trap levels each with their own independent quasi-Fermi-level. This is shown in Figure~\ref{fig:figureX}, where the Shockley--Read--Hall (SRH) capture escape equations \cite{shockley1952statistics} are solved for each trap, enforcing detailed balance of charge carriers. For an electron trap the following equation is solved,
\begin{equation}
    \frac{\partial n_t}{\partial t} = r_{ec} - r_{ee} - r_{hc} + r_{he},
\end{equation}
where $r_{ec}$ is the rate of electron capture, $r_{ee}$ is the rate of electron escape, $r_{hc}$ is the rate of hole capture, and $r_{he}$ is the rate of hole escape. The former two terms represent carrier trapping, the latter two represent recombination.

An analogous set of equations is solved for hole traps. Fermi--Dirac statistics are used for trapped carriers. The charge distribution of trap states $\rho$ is described with an exponential density-of-states that is given in Equation \ref{equ:tail}. $E$ is the energetic distance from the transport energy and $N_0$ is the density of traps at the transport energy.

\begin{equation}\label{equ:tail}
    \rho(E)=N_0 e^{-{{qE} \over {kT}}}.
\end{equation}

This approach allows us to describe the carrier density in terms of both, energy and position space across the three-dimensional structure. Figure~\ref{fig:figure1}d shows the trapped carrier density over the device, while Figure~\ref{fig:figure1}e plots the average conductivity including trap states.

\section{Numerical challenges}

It should be noted at this point that 3D drift--diffusion models are notoriously difficult to solve\cite{scharfetter2005large,selberherr1984analysis, bank1983numerical, jerome2012analysis}. The difficulty lies in the non-linearity of the problem, and that the physical system contains very small numbers and very large numbers: As the system of equations grow, the small numbers get lost in the larger numbers (round-off errors), and the system of equations become unstable.  Furthermore, memory/compute time grows by a factor of at least $n^3$ as the number of mesh points ($n$) increases in each dimension. Practically, this means systems of $20\times20\times20$ equations are easy to solve on a laptop within seconds, systems of $30\times30\times30$ equations needs a work station and a compute time approaching a hour, and systems of $50\times50\times50$ become extremely challenging, often requiring access to high-performance computing resources. Beyond this size, numerical instability, memory limitations, and convergence failures become dominant barriers preventing simulations of this size being used for routine analysis. This is explained in more detail in the SI, but for now one should consider the ability to solve $50\times50\times50$ drift--diffusion equations for a well-defined MOSFET-type structure as the state-of-the-art.

Unfortunately, BHJ structures in organics are more complex than typical inorganic devices tackled by commercial solvers, because they contain a complex structure that can change from simulation run to simulation run. This makes it harder to optimize a solver for a given problem. Furthermore, mobilities often vary over many orders of magnitude which can exacerbate numerical difficulties when one comes to solve the equations. Standard drift--diffusion models have 3 variables per mesh point ($\phi$,$F_n^f$,$F_p^f$), however, models for disordered organic semiconductors commonly have 5 trap states per mesh point for each electron/hole population: This means one has to solve for ($\phi$,$F^f_n$,$F^f_p$,$5 \times F^t_n$, $5 \times F^f_p$) variables per mesh point. All these factors combined puts the problem on the brink of what can be solved with modern computers.

To overcome this we tested out many different strategies, this journey is detailed in full in the SI. However, the strategy that worked best in terms of results, performance and stability was to solve the 3D problem in uncoupled 2D slices normal to the substrate, then rotate the solver through 90 degrees, and again solve uncoupled 2D slices, again normal to the substrate but in the other direction. Thus we ended up with two quasi-3D simulations each of which accounted current flow only in one lateral direction.  We then combined current flows/carrier densities from the two simulations to obtain our final results. This worked well, because the strongest current flowed normal to the substrate (due to the field) -- our solver captured this flow fully, however it also enabled us to capture the weaker current flows in both directions normal to the substrate. Decoupling approaches like this have a long history in semiconductor device simulators \cite{selberherr1984analysis, lim2009design}.

\section{Results}

\begin{figure}
    \centering
    \includegraphics[width=\columnwidth]{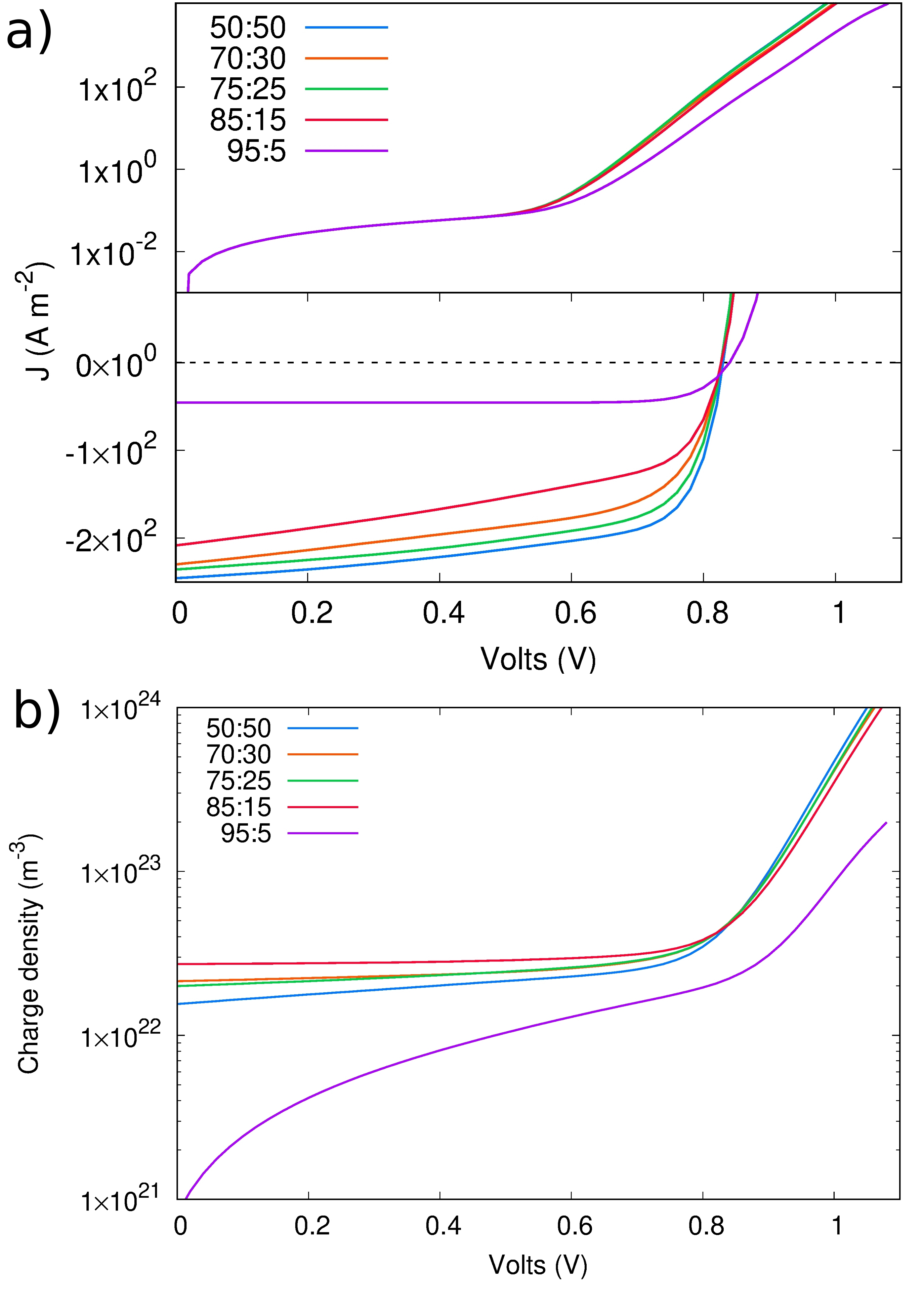}
    \caption{a) Light and dark JV curves for morphologies with blend ratios between 50:50--95:5. It can be seen that -- as would be experimentally be observed -- the shape of the JV curves are strongly dependent upon the morphological configuration. b) Charge carrier density as a function of voltage under AM1.5G illumination.}
    \label{fig:figure_jv}
\end{figure}

Figure~\ref{fig:figure_jv} presents the simulated light and dark JV curves for morphologies with donor--acceptor ratios between 50:50--5:95. The dark JV curves are shown on the top of the plot. It can be seen that by changing the morphology, the slope of the dark JV curve changes in the diode region indicating a change of recombination rate with carrier density. The shunt portion of the curve (below 0.5~V) is modeled with an external resistor so does not change. The light JV curves are plotted on the bottom of the figure, and it can be seen that the blend ratio significantly affects the solar cell parameters $\jsc$, $\Voc$ and $FF$. The device with the highest photocurrent is the 50:50 blend, while the device with the lowest is the 95:5 blend ratio. This result is expected because more equal blend ratios will reduce the average distance from a photogeneration site to the interface and encourage efficient free charge generation.

Figure~\ref{fig:figure_jv}b shows the charge carrier density as a function of voltage under AM1.5G illumination. It can be seen that the general trend is the same as measured experimentally through charge extraction measurements \cite{Mackenzie2012extractingmicroscopic}. That is, a relatively gentle increase in charge density until just below $\Voc$, followed by a more steep increase. It is worth noting that the magnitudes of the curves are close to those previously reported \cite{shuttle2010charge} at around $10^{22}$--$10^{23}~$~m$^{-3}$. The device with the lowest charge density is that of the 95:5 blend. This would be expected, as there are few interfaces for exciton dissociation and charge generation. It is interesting to note that the most efficient device with a ratio of 50:50 has a slightly lower charge density than all devices except the 95:5 blend, which is caused by the better extraction due to the highly-interconnected morphology.

\begin{figure}
    \centering
    \includegraphics[width=\columnwidth]{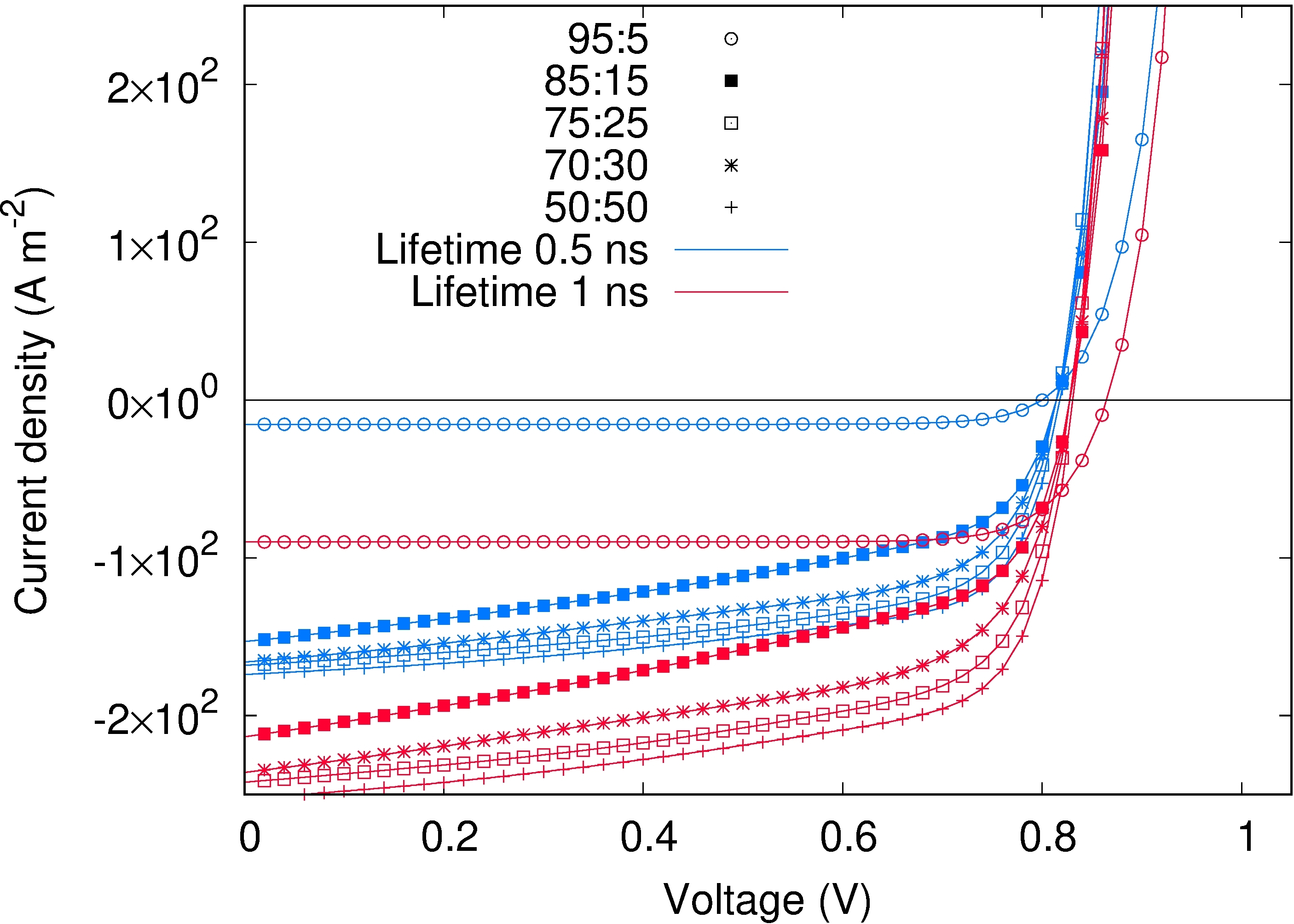}
    \caption{JV curves calculated for different radiative loss time constants (\( \tau_{PL} = 1/k_{PL} \)) plotted for each of the morphologies under test. It can be seen that while the chosen morphology is important for determining overall device performance, microscopic parameters which are determined by microscopic material parameters are equally as important.}
    \label{fig:lifetime}
\end{figure}

Figure~\ref{fig:lifetime} presents two sets of JV curves for each morphology, where the exciton radiative loss time constant (\( \tau_{PL} = 1/k_{PL} \)) was varied between 1~ns (red lines) and 0.5~ns (blue lines). Radiative losses typically dominate exciton recombination away from interfaces under steady-state operating conditions (i.e., not in high-power pulsed regimes where bimolecular annihilation, \( \alpha X^2 \), can dominate). The value of \( \tau_{PL} \) is primarily dictated by microscopic material parameters. It can be seen that increasing \( \tau_{\mathrm{PL}} \) by only a factor of two (from 0.5~ns to 1~ns) leads to a significant improvement in device performance. In fact, the performance shift caused by slightly changing the lifetime is comparable to that observed when switching between entirely different macroscopic morphologies. This suggests that while the large-scale morphology plays an important role in device performance, microscopic material properties are equally if not more important.

\begin{figure}
    \centering
    \includegraphics[width=\columnwidth]{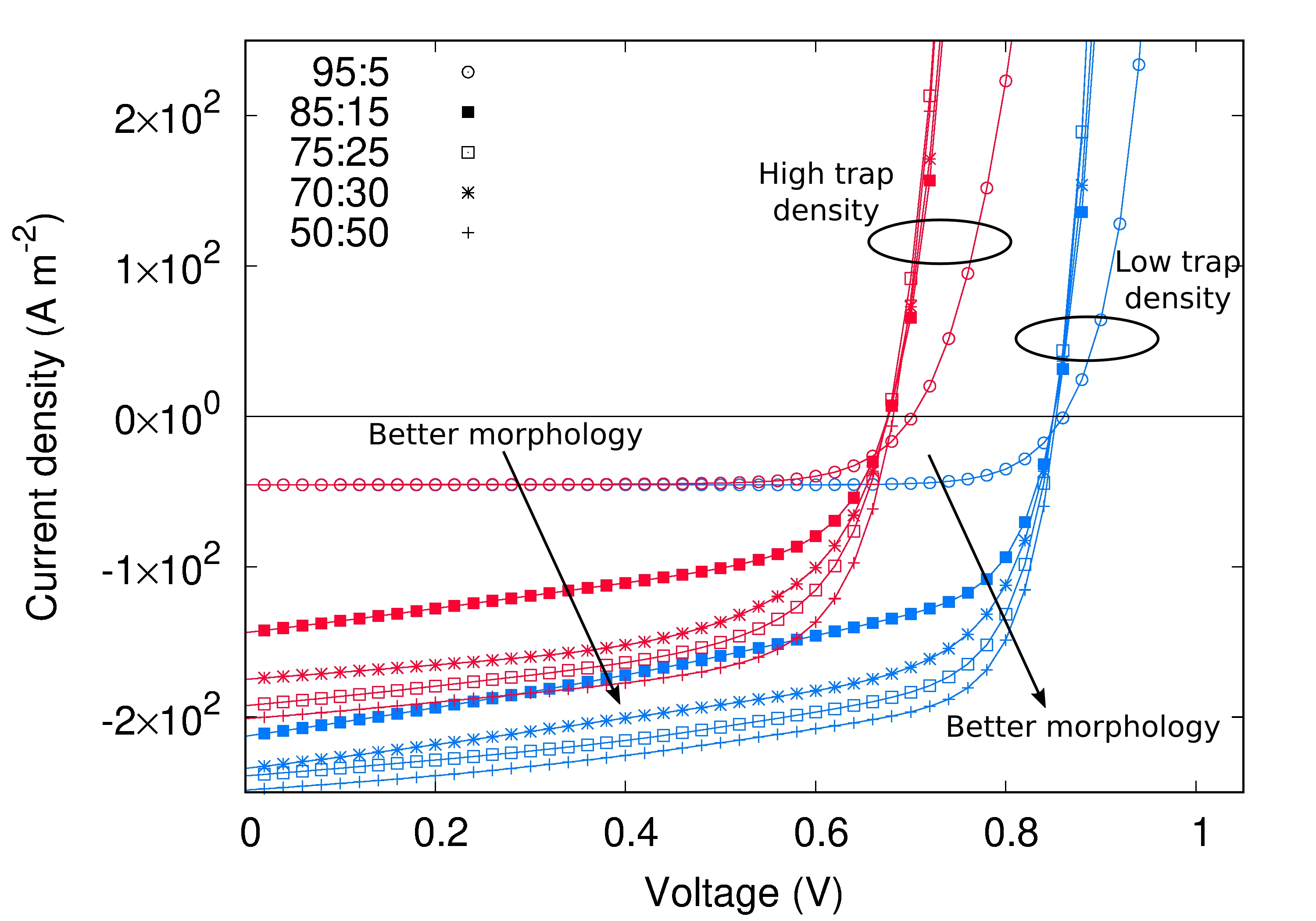}
    \caption{Different blend ratios plotted for two different microscopic trap densities of $5 \times 10^{23}$ and $1 \times 10^{26}$~m$^{-3}$. It can be seen that both nanomorphology and energetic disorder play an important role in achieving a high efficiency.}
    \label{fig:trap_density}
\end{figure}

Two sets of JV curves are shown in Figure~\ref{fig:trap_density}, one with a high trap density of $N_0 = 1 \times 10^{26}$~m$^{-3}$ and the other with a low trap density of $N_0 = 5 \times 10^{23}$~m$^{-3}$. The device with a high trap density is representative of what might be found in a P3HT:PCBM device, while the low density corresponds to a state-of-the-art high-performance device. In both cases, as the blend ratio approaches 50:50, it can be seen that the shape of the JV curve improves. The second visible trend is that the devices with a lower density of trap states generally outperform all devices with a higher trap density. From this, we can conclude to achieve maximum efficiency the device must have a well-connected nanomorphology with low energetic disorder.  

\begin{figure}
    \centering
    \includegraphics[width=\columnwidth]{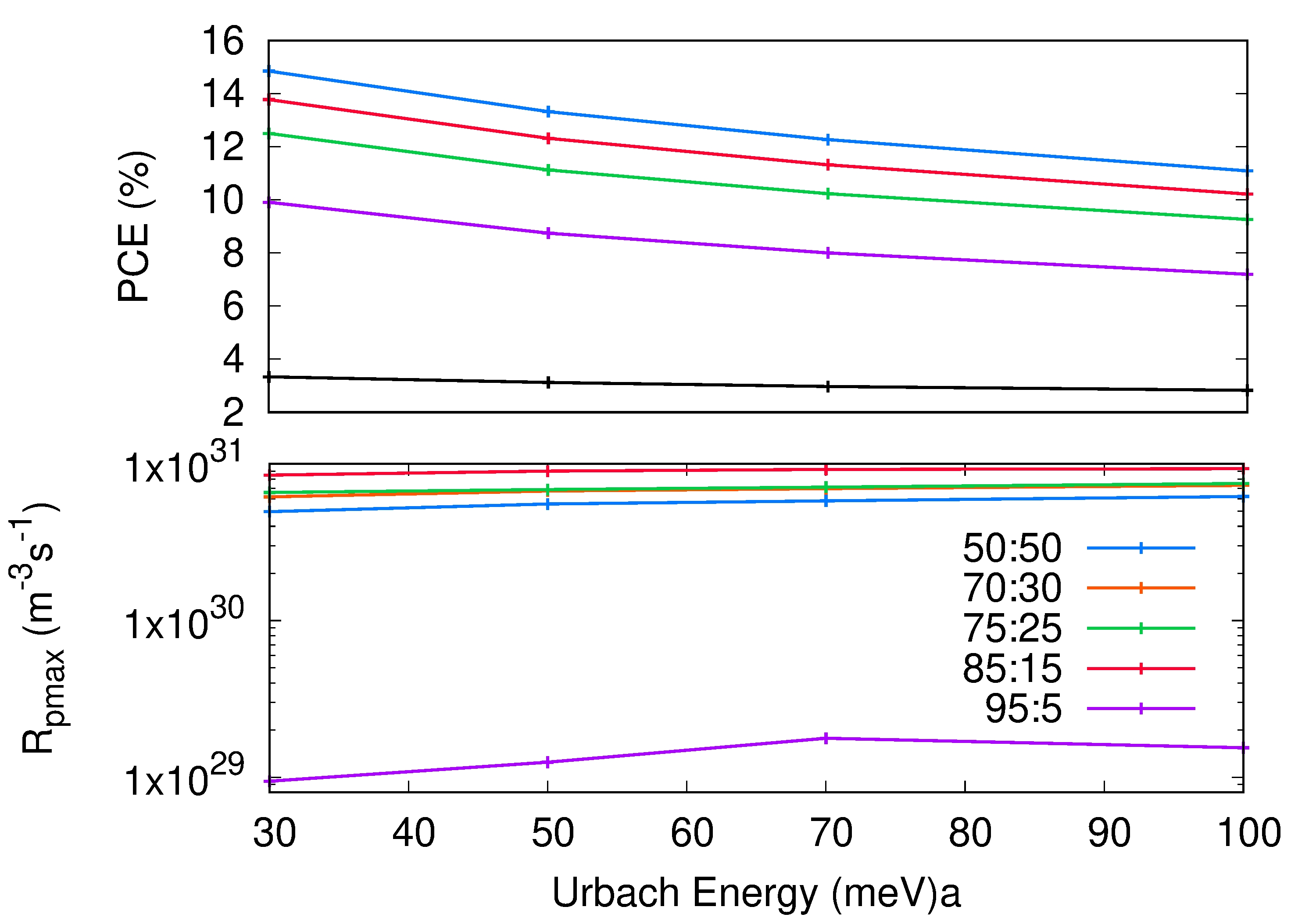}
    \caption{Top: Power conversion efficiency against Urbach energy. It can be seen in this example that the nanomorphology has a larger impact than the energetic disorder. Bottom: Short circuit current density $\jsc$ plotted against Urbach energy.}
    \label{fig:figure_pce_R_pmax}
\end{figure}

Figure~\ref{fig:figure_pce_R_pmax} shows (top) the power conversion efficiency (PCE) and (bottom) the recombination rate at the maximum power point as a function of the characteristic energy of the exponential trap density -- also called Urbach tail slope energy -- for a device with a trap density of $1 \times 10^{24}$~m$^{-3}$. This concentration of trap states is somewhere between the two curves in Figure~\ref{fig:trap_density} and would be considered a fairly good device by modern standards. It can be seen in this example, that the highest efficiency devices all have low levels of energetic disorder, but overall the nanomorphology has a greater impact. The exact impact that energetic disorder will have on a system will be affected by Urbach tail slopes, absolute trap density, as well as the carrier recombination and trapping cross sections. These effects will be convolved with the spatial overlap of the electron/hole populations which will to a large degree be determined by the nanomorphology. Thus, it is hard to say one parameter is key for device performance as they are all coupled non-linearly, but one can say that all parameters, both microscopic and macroscopic, must be optimized for an efficient device. In device design, this means that molecules that do not pack well due to unnecessarily long aliphatic side chains, potentially leading to higher trap densities, will not be as high-performing even if the nanomorphology is favorable. 

The bottom panel of Figure~\ref{fig:figure_pce_R_pmax} presents the recombination rate as a function of Urbach tail slope. One would naively expect best performing devices to have the lowest recombination rate, and the worst the highest. However, the trend is more complex: The 95:5 device has the lowest recombination rate, followed by the 50:50 device, while the other devices have higher rates. This can be explained by the 95:5 device not being able to generate charge carriers efficiently, leading also to a lower recombination rate; whereas the 50:50 device, being able to generate charge carriers most efficiently, has an efficient extraction due to favourable transport pathways for both charge carrier types. The other devices show a mixture of both effects. This highlights the point that the exact performance is a combination of microscopic and macroscopic parameters.

\begin{figure*}
    \centering
    \includegraphics[width=\textwidth]{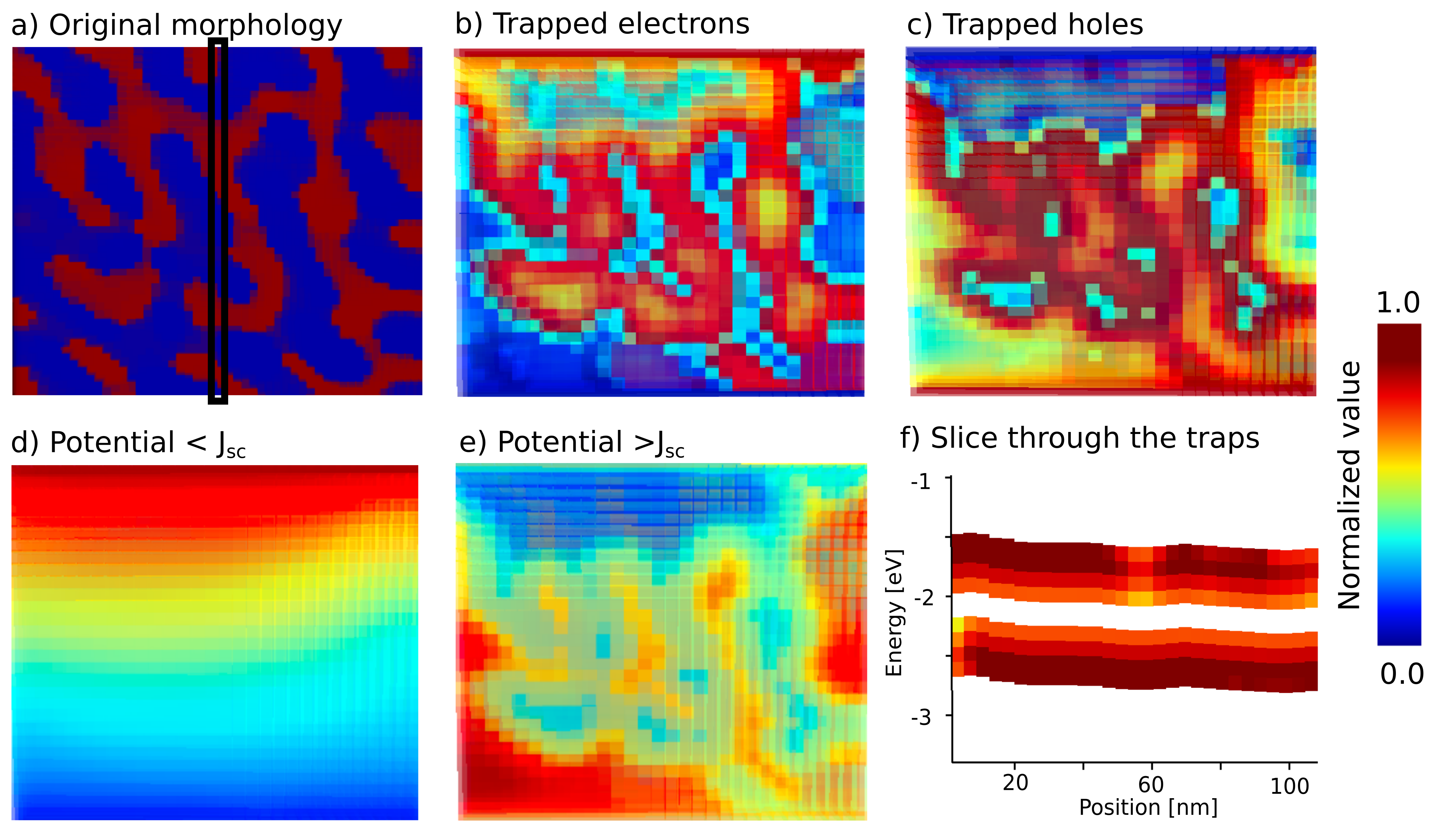}
    \caption{a) A slice down the 3D morphology; b) Trapped electron density; c) Trapped hole density; d) A map of potential across the device below $\Voc$; e) A map of potential across the device above $\Voc$; f) A slice through the electron and hole transport energies and occupied trap levels of the device at high injection current.}
    \label{fig:figure_bands}
\end{figure*}

Figure~\ref{fig:figure_bands}a plots a slice down a 3D morphology from top contact to bottom contact, while Figures~\ref{fig:figure_bands}b and c show the spatially resolved trapped electron and trapped hole densities. The trapped electrons mostly reside in the (red) acceptor, while the holes mainly reside in the (blue) donor material. The interlocking fingers of the BHJ can be clearly seen to spatially separate the electron and hole populations. Figure~\ref{fig:figure_bands}d presents the LUMO transport energy as a function of position at around 0.45~V (below $\FF$ and $\Voc$); it can be seen that the potential gradually changes from the top to bottom of the device. However, it is not uniform due to the trapped charge carriers. Thus, a standard assumption that charge carriers within the device experience a potential that is merely a function of voltage is not true. Figure~\ref{fig:figure_bands}e shows the potential distribution above $\Voc$. Here, at higher carrier densities, islands of charge carriers start to emerge in the BHJ and the potential can no longer be considered uniform across the device. Figure~\ref{fig:figure_bands}f, in turn, shows the distribution of trapped carriers in energy space and position space down the center of Figure~\ref{fig:figure_bands}a (black box) above $\Voc$. Both potential and carrier distributions change significantly as a function of space and energy.

\begin{figure}
    \centering
    \includegraphics[width=\columnwidth]{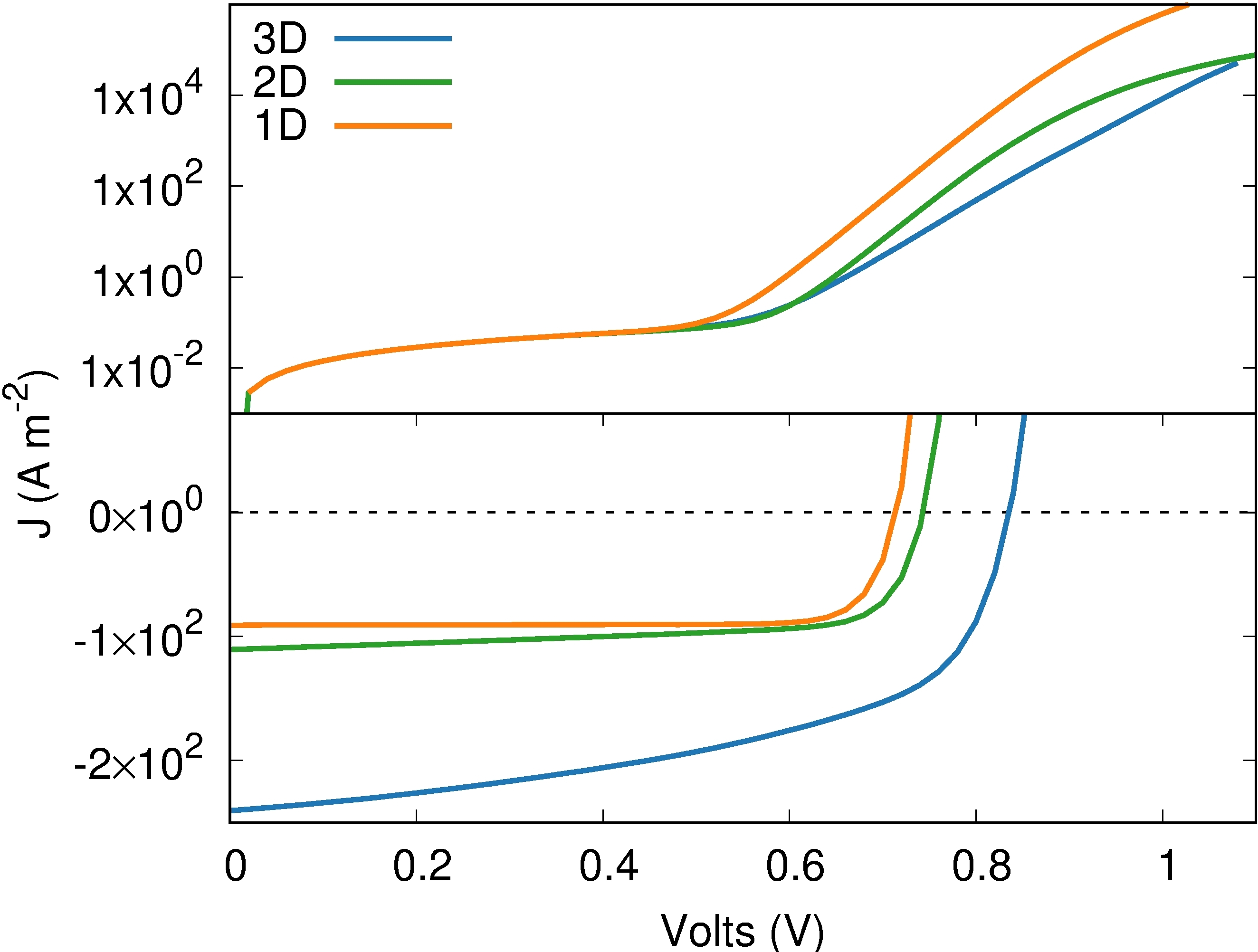}
    \caption{Comparison of 3D, 2D, and 1D models. It can be seen that both the 1D effective medium model and the 2D slice of the BHJ underestimate $\Voc$ and $\jsc$. This is because in the 1D effective medium models, there is perfect spatial overlap between electrons and holes, the percolation paths are limited in the 2D model, while the 3D model captures both of these effects.}
    \label{fig:figure5}
\end{figure}

Figure~\ref{fig:figure_bands} demonstrates that both the distribution of trapped charges and the band structure inside the BHJ is far removed from the simple pictures presented by 1D models. We compared how different the results are from more simple 1D models: We kept the same device parameters, but ran the simulation firstly in 2D -- while preserving the morphology map from the 3D model --  and in 1D, where an effective medium approach was used. The results are shown in Figure~\ref{fig:figure5}. If one examines the top panel, it can be seen that going from the 1D effective medium model to 2D to 3D models, the slope of the dark curve reduces, indicating less recombination and current flowing through the device at any given voltage. This is because the 1D effective medium model assumes perfect spatial overlap of electrons and holes enabling them to easily recombine. Although the 2D model spatially separates electrons and holes, it cannot capture current flow through the 3D structure or around obstacles. The bottom panel shows that $\Voc$, $FF$ and $\jsc$ increase as the dimensionality of the models increases. This is for the same reasons as given above, mainly that the 3D model has more current paths through the medium and spatially separates the electron and hole distributions. As above, the shunt resistance is modeled as a simple resistor in parallel with the device, hence the shunt region of the dark curve does not change between simulations.

\section{Conclusion}
In this work, we presented a hybrid simulation framework that bridges the gap between full KMC and traditional drift--diffusion models by incorporating realistic three-dimensional BHJ morphologies and a detailed treatment of trap states. By combining phase-field morphology generation with a quasi-3D drift--diffusion solver including an exponential tail of localized traps, we captured both configurational and energetic disorder in organic solar cells. Our results show that even well-connected nanomorphologies can underperform due to energetic disorder, especially at high carrier densities. This highlights the need to look beyond nanomorphological optimization and instead focus on controlling molecular-level packing during material design and processing. By varying the Urbach energy, we found that both morphology and energetic disorder impact performance, with macroscopic connectivity often playing the dominant role. Nonetheless, reducing trap densities through improved molecular packing remains critical. We also demonstrated that simplified 1D and 2D models significantly underestimate device performance due to their inability to capture spatial charge separation and 3D percolation pathways. Full 3D modeling is therefore essential for accurate predictions of $\Voc$, $\jsc$, and fill factor. Achieving high-efficiency organic photovoltaics requires simultaneous control of both morphology and energetic disorder. The presented framework offers a powerful tool for disentangling these effects and guiding future materials and device design.

\section{Acknowledgements}
We thank the Deutsche Forschungsgemeinschaft (DFG) for funding this work (Research Unit FOR 5387 POPULAR, project no.\ 461909888).

\FloatBarrier
\bibliography{main}

\appendix
\newpage
\onecolumngrid

\section{Supporting Information}\label{sec:SI}

\subsection{Morphology generation}\label{ssec:SI_morphology}
\subsubsection{Model equations}
The morphology of a BHJ is formed by the drying of a thin film. A dilute blend containing polymer, NFA, and solvent is distributed on a substrate. As the solvent evaporates, the mixture undergoes phase separation into polymer rich and NFA rich areas. Describing this process in detail results in a large, nonlinear, coupled system. For instance, the solvent concentration influences the rate of donor-acceptor separation, while the mixtures composition, in turn, affects the evaporation rate. In addition, there are hydrodynamic effects having an influence on the fluid composition. Taking all effects into account leads to a numerically expensive model, not suited to generate the large number of morphologies needed as input for the charge transport model. To this end, we revert to a simplified phase-field model -- based on the more complex framework presented in \cite{PCiloglu_2025} -- which still produces realistic morphologies.

Introducing the volume fractions of polymer, NFA, and solvent, respectively, 
\begin{equation}
    \phi_p, \phi_{nfa}, \phi_s : \Omega \times [0,T] \to [0,1], 
\end{equation}
we can describe the morphology evulation in the domain $\Omega \subset \mathbb{R}^3$ over time $[0,T]$. Using the conservation relation $ \phi_s = 1-\phi_p-\phi_{nfa}$, the morphology description is reduced to the evolution of the polymer and NFA volume fractions. Both phase separation and evaporation are driven by the minimizations of the free energy of the system. To keep the model manageable, evaporation of solvent is modelled by an outward flux of solvent, with a corresponding inlflux of polymer and NFA, at the top of the domain, see, e.g., the approach in \cite{KBergermann_2023}. This not only simplifies the description of the evaporation process, but also allows for the omission of the air-phase needed in the free energy description presented in \cite{PCiloglu_2025}, reducing the size of the discrete system significantly. Phase separation is still modelled by the minimization of the free energy functional $F = \int_\Omega f \ dx $. The free energy density change $f = f^{loc} + f^{nonloc}$ is driving the phase separation and consists of a local part describing the change in mixing energy and a nonlocal part penalizing the field gradients. To desscribe the change in local mixing energy density, we make use of a Flory–Huggins-type potential of the form 
\begin{equation}
    f^{loc}(\boldsymbol{\phi})  = \frac{RT}{V_0}\left( \sum_i \frac{\phi_i}{N_i} \ln \phi_i + \sum_i \sum_{j<i} \chi_{i,j} \phi_i \phi_j \right) 
\end{equation}
where $N_i $ describes the molar size and $\chi_{i,j}$ the Flory-Huggins interaction parameter. The constant parameters $V_0$, $R$, and $T$ represent the reference molar volume, gas constant, and temperature, respectively. The nonlocal term penalizing the field gradients has the form
\begin{equation}
    f^{nonloc}(\boldsymbol{\phi})  = \sum_i \frac{\epsilon_i^2}{2} \left( \nabla \phi_i\right)^2 .
\end{equation}

Driving phase separation by minimizing the free energy yields the governing Cahn-Hilliard equations, 
\begin{equation}\label{eq:AppCH}
    \frac{\partial \phi_i }{\partial t } = \nabla \cdot ( M_i \nabla \mu_i) \quad \text{ and } \quad 
    \mu_i = \frac{\partial f^{loc}}{\partial \phi_i} - \epsilon_i^2 \nabla^2 \phi_i \text{ in } \Omega \times [0,T] ,
\end{equation}
with $i \in \{p, nfa \}$. Here, $\mu_i$ is the chemical potential, $M_i$ the mobility coefficients, and $\epsilon_i$ the surface tension coefficients. Adding the boundary conditions
\begin{equation}
    \nabla \mu_i \cdot \bm{n} = \begin{cases}
    -k\phi_p \phi_s, &\text{ on } \Gamma_t \times [0,T] \\
    0, &\text{ on } \partial \Omega \setminus \Gamma_t \times [0,T]
    \end{cases}
\end{equation}
and initial conditions
\begin{equation}
    \phi_p(x, 0) = \phi_p^0(x) \text{ in } \Omega \quad \text{ and } \quad \phi_{nfa} (x, 0) = \phi_{nfa}^0 (x) \text{ in } \Omega
\end{equation}
to the model completes the system of governing equations. The flux boundary conditions defined at the top of the domain, $\Gamma_t \times [0,T]$, represent the evaporation of the solvent at rate of the evaporation proportionality constant $k > 0$. Using suitable scalings of the above equations leads to the desired Equation~\eqref{eq:CH}.

\subsubsection{Discretization and Preconditioning}
To solve the governing equations numerically, we employ a finite element approach with a semi-implicit discretization in time, where the linear parts of the right-hand side of ~\eqref{eq:CH} are treated implicitly, and the nonlinear free energy terms are handled explicitly. This results in a system of the form

\begin{equation}\label{eq:CH_sys}
    \frac{1}{\tau} \bm{M}\left( \bm \phi_i^{(l+1)} - \bm \phi_i^{(l)} \right) = - \gamma_i\bm K \bm \mu_i ^{(l+1)} \quad \text{ and } \quad \bm M \bm \mu_i^{(l+1)} = \bm f_i^{(l)} + \beta_i \bm K \bm  \phi_i^{(l+1)} 
\end{equation}
at each time $t^{(l)} = l \tau $. Here, $\bm M $ and $\bm K$ denote the standard mass and stiffness matrices arising from the spatial discretization, and $\bm \phi_i$, $\bm \mu_i$, and $\bm f_i$ are the coefficient vectors of the discretized phase-field parameters, chemical potentials, and discretized representations of $\frac{\partial f}{\partial \phi_i}$, respectively. Because these equations must be solved repeatedly - due to typically small time step sizes $\tau$ in explicit schemes and a large number of mesh points in three-dimensional domains- preconditioning techniques are essential to accelerate the computations. Following the block-diagonal preconditioning strategy presented in~\cite{KBergermann_2023, PCiloglu_2025}, we solve the equivalent linear system

\begin{equation}\label{eqn:CH_linear}
        \begin{bmatrix}
            \mathbf{M}  &  \tau\gamma_i \mathbf{K}  \\
            \tau\gamma_i\mathbf{K} & - \frac{\tau\gamma_i}{\beta_i}\mathbf{M} 
        \end{bmatrix}  
        \begin{bmatrix}
          \boldsymbol{\phi}_i^{l+1}  \\
          \boldsymbol{\mu}_i^{l+1}
        \end{bmatrix} = 
        \begin{bmatrix}
           \mathbf{M} \boldsymbol{\phi}_i^{l} \\ 
           -\frac{\tau\gamma_i}{\beta_i}\boldsymbol{f}_i^{l}
        \end{bmatrix},  
\end{equation}
which forms a symmetric saddle-point system. The suitable preconditioner is chosen as
    \begin{align*}
       P= \begin{bmatrix}
            \mathbf{M} & 0 \\
             0 & \mathbf{S}
        \end{bmatrix} ,        
    \end{align*} 
where $ \mathbf{S} =\frac{\tau\gamma_i}{\beta_i} \mathbf{M} + \left( \tau\gamma_i \right)^2 \mathbf{K}\mathbf{M} ^{-1} \mathbf{K} $  is the (negative) Schur complement.  The model \eqref{eq:CH_sys} is solved with a \texttt{PYTHON} implementation using the finite element libraries \texttt{DOLFINX} \cite{dolfinx}, \texttt{Basix} \cite{basix}, and \texttt{UFL}  \cite{ufl} from the \texttt{FENICS} project  \cite{fenics, autofenics} with the version 0.8.0. We apply the preconditioned \texttt{MINRES} method~\cite{saad2003iterative} as the Krylov subspace solver, with algebraic multigrid (\texttt{AMG}) preconditioning using the Ruge–Stüben method.

Letting the model evolve over time until it reaches a quasi-stationary state, the morphology shows well-formed phase-separated domains. The initial conditions, including small random perturbations, are chosen to match experimentally relevant donor--acceptor blend ratios, allowing us to test device performance across a range of morphologies. The resulting morphologies are then used as static input in the charge transport simulations described in Section~\ref{sec:ElecMod}.

\subsection{3D drift--diffusion modeling}
This section provides a commentary on the technical challenges encountered and the journey we went on when we decided to start solving large systems of drift--diffusion equations to describe bulk heterojunction (BHJ) devices in three dimensions. We highlight key computational difficulties and describe the strategies employed in this work to overcome them. This is not meant to be an exhaustive description of the problem, it is however meant to be useful to others who come after us.

\subsubsection{An overview of the problems}
Solving three-dimensional drift--diffusion problems in BHJ organic semiconductor devices presents significant computational challenges, even with modern semiconductor models on state-of-the-art hardware. At the heart of solving the drift-diffusion problem is the Newton-Raphson iteration, where the nonlinear drift--diffusion-Poisson system is linearized to form a large system of the form:
\begin{equation}
\mathbf{J} \, \delta \mathbf{x} = -\mathbf{F}(\mathbf{x}),
\label{eq:newton}
\end{equation}
where $\mathbf{x}$ represents a vector containing the variables one wants to solve for. In our case this is potential, and Fermi levels throughout our finite difference grid.  $\mathbf{F}(\mathbf{x})$ is an error function, that describes how far each variable is from the correct answer. It is formed by setting each equation at each mesh point to an error function $f$ (i.e. Poisson's equation at point mesh point 0 would form one of these equations):
\begin{equation}
\nabla \cdot [\epsilon(\mathbf{r_0}) \nabla \phi(\mathbf{r_0})] +q [p_{f/t}(\mathbf{r_0}) - n_{f/t}(\mathbf{r_0})]=f_0(\phi,\mathbf{r},..).
\end{equation}

The electron continuity equation at the 0th mesh point would be another one of these equations: 
\begin{equation}
\nabla \cdot \mathbf{J}_n(\mathbf{r_0}) - q \left[ G(\mathbf{r_0}) + R(\mathbf{r_0}) \right]=f_1(\phi,\mathbf{r_0},..).
\end{equation}

By writing the entire equation set one is interested in as error functions, one can describe how far the set of equations is to being perfectly solved. If the system were perfectly solved the sum of all the vector $\mathbf{F}$ would be zero. In practice, we can never solve it perfectly due to numerical errors but we can get close.

The Jacobian matrix denoted by $\mathbf{J}$ represents the derivatives of the error functions at each point in the grid. This matrix has many off-diagonal elements coupling the various equations within the problem. To solve for \( \mathbf{x} \), one must either directly or iteratively compute a solution to the system defined by \( J \).

Once equation \ref{eq:newton} is solved one obtains an update, $\delta \mathbf{x}$, to the variables one is interested in. One then simply adds the update vector ($\delta \mathbf{x}$) to the present values of $\mathbf{x}$, recalculates the error functions and Jacobain and solves the system again and again until the sum of $\mathbf{F}$ falls to an acceptable level. For problems with accurate derivatives in $\mathbf{J}$, error should reduce super linearly and the whole system of equations should solve in around 10 steps.

For a simple one-dimensional problem, the Jacobian matrix $J$ is typically narrow-banded and highly sparse, making it efficient to solve using direct or iterative methods. As the dimensionality of the system increases, the number of non-zero off-diagonal elements grows, and the Jacobian becomes less sparse. This increases both the computational and memory burden when solving the associated linear system, regardless of the solver used (e.g., LU, Cholesky, or Krylov methods).

In a three-dimensional discretized domain with \( n \) points per dimension, \( J \) becomes an \( n^3 \times n^3 \) matrix. While the Jacobian itself remains sparse, any direct inversion or even LU factorization can lead to substantial fill-in, making storage and computation scale poorly. For instance, a \( 10 \times 10 \times 10 \) system is manageable, but a \( 20 \times 20 \times 20 \) problem already stresses standard hardware, and a \( 50 \times 50 \times 50 \) system---approaching 125{,}000 unknowns---often becomes intractable without specialized solvers or high-performance computing resources. 

Compounding these challenges is the extremely wide dynamic range of values in the drift--diffusion Jacobian, which can span up to 20 orders of magnitude due to high generation rates and strong local electric fields. Many physical quantities that need to be considered often span many orders of magnitude. For example carrier concentrations in trap states can range from $1\times10^{-6}$ to $1\times10^{25} m^{-3}$.  Given that the numerical values span such a vast range of scales, the solvers can be prone to numerical round-off errors. This is especially important when two large numbers of approximately similar magnitude are subtracted and one often loses significant digits in the corresponding numerical result.  Thus being able to solve the set of equations is not simply a matter of having more powerful hardware.

\subsubsection*{Computational Challenges in BHJ Systems}

\textbf{Trap States in Organic Semiconductors:} Organic systems introduce additional complexity because each mesh point must include multiple quasi-Fermi levels to represent localized electron and hole traps. For example, a single spatial node may require 5--10 trap levels for electrons and another 5--10 for holes, dramatically expanding the dimensionality of the unknown vector \( \delta \mathbf{x} \) and the Jacobian matrix \( \mathbf{J} \). This increase not only places greater demands on memory but also significantly slows convergence in Newton-based solvers, due to the stiff and highly nonlinear nature of trap-related dynamics. By contrast, conventional semiconductor models typically require only one Fermi level per carrier type at each mesh point. For BHJ solvers that also include singlets/tripplets, this adds more unknowns and makes the problem yet harder. 

\vspace{1em}
\textbf{Randomness in BHJ Morphologies:} Unlike traditional semiconductor devices (e.g., MOSFETs), BHJ systems exhibit strong spatial randomness due to the intermixed donor--acceptor domains and intrinsic structural disorder. As a result, each morphology defines a unique configuration of material interfaces, phase connectivity, and energetic landscape. Consequently, each simulation becomes a distinct numerical problem, limiting the effectiveness of solver preconditioning and reuse. This variability poses a significant challenge to the robustness, generality, and scalability of conventional numerical methods.

\vspace{1em}
\textbf{Large, Unpredictable Dynamic Range of Variables:} Organic semiconductors with trap states exhibit a significantly broader range of carrier densities than standard semiconductor models. This arises because the occupation of trap states can vary from nearly empty to nearly full, even within the energy range associated with a single mesh point. As a result, the system spans many orders of magnitude in local carrier concentrations, introducing numerical stiffness and increasing the risk of instability in both the residual evaluation and the Jacobian matrix.

\subsubsection*{Approaches to Overcoming These Challenges}

To address the computational difficulties outlined above, we employed a range of techniques drawn from both standard drift--diffusion modeling and custom strategies tailored to the peculiarities of BHJ systems.

\vspace{1em}
\textbf{Standard Techniques:} As a baseline, we adopted several well-established methods commonly used in semiconductor drift--diffusion simulations. Notably, we used the Scharfetter–Gummel discretization scheme, which solves a boundary-layer problem between adjacent mesh points to more accurately compute current densities. This approach is essential because carrier densities vary exponentially with the electric field, and naive central differences can result in large numerical errors or instabilities. 

In addition, we carefully implemented the Bernoulli function to preserve numerical accuracy in the low-field and high-field limits, where floating-point cancellation or overflow can occur. These standard techniques are essential to ensure correct treatment of transport physics, especially under conditions of steep gradients or large trap populations.

\vspace{1em}
\textbf{ADI Scanning:} We explored a range of coupled and decoupled solution strategies, including the use of \emph{Alternating Direction Implicit} (ADI) scanning methods. In this approach, the three-dimensional problem is decomposed into slices, which are sequentially updated along one spatial direction at a time. For example, we performed forward sweeps along the \( x \)-axis (from \( x = 0 \) to \( x = x_{\text{max}} \)), followed by similar sweeps along the \( z \)-axis, in an effort to propagate updates efficiently across the domain.

We experimented with a variety of coupling schemes between these slices, including fully coupled block updates and partially coupled configurations -- for instance, solving groups of four \( x \)-slices coupled with three mesh layers in the \( z \)-direction. Overlapping blocks and alternating sweep directions were also tested to enhance stability and convergence.

However, as the problem size increased -- particularly for grids exceeding 40 nodes in a given direction -- convergence became increasingly difficult. The long physical distance between opposing boundaries meant that information introduced at one end required many iterations to affect the far side of the domain. This led to very slow convergence and, in some cases, ambiguity as to whether the solver was progressing at all.

We also tested simplified schemes such as uncoupled vertical column updates and partial decoupling of the equations -- for example, solving the Poisson equation separately from the continuity equations for electrons and holes. While such decomposition can reduce matrix bandwidth and inter-variable coupling, it proved ineffective in our case. The strong nonlinearities introduced by trap states in both electron and hole transport equations reintroduced significant coupling, preventing successful decoupling.

To support rapid experimentation, we implemented the ADI scheme and its variants within a Lua scripting environment embeded into OghmaNano, allowing users to configure sweep directions, block sizes, coupling strategies, and update order dynamically. This flexibility enabled rapid testing of solver configurations and parameter tuning. This is described fully in the OghmaNano manual which can be found on the webpage https://www.oghma-nano.com.

\vspace{1em}
\textbf{Solver Strategies:} We experimented with a variety of solvers to handle the large linear systems arising within the Newton iterations. Initially, we used UMFPACK, which offered good performance for smaller problems due to its efficient sparse LU factorization. However, as the problem size increased, UMFPACK became impractical due to memory limitations and its lack of parallelism -- it is strictly single-threaded and thus inherently limited on modern multicore hardware.

To address these limitations, we tested alternative sparse direct solvers, including MUMPS \cite{amestoy2000mumps} and SuperLU \cite{li2005overview}. Unfortunately, we encountered stability and scalability issues with these solvers in our application context. Ultimately, we adopted the PETSc library \cite{petsc-user-ref}, which provided access to a wide range of iterative and direct solvers, flexible matrix formats, and built-in MPI support.

To integrate PETSc into our simulation workflow, we employed a hybrid architecture. Our main application remained single-threaded, but we interfaced with PETSc solvers running in MPI mode via a shared-memory communication layer. This was achieved using memory-mapped files (\texttt{mmap}) to expose shared buffers, coordinated by POSIX semaphores to ensure thread-safe data exchange. This architecture allowed us to leverage fully distributed PETSc solvers without requiring the entire application to be MPI-aware.

For the linear system solution, we used a distributed-memory solver stack built on PETSc and MPI. The global Jacobian matrix is partitioned across MPI ranks using a block-row distribution, with each process owning a subset of rows. Matrix assembly and residual vector population are performed in parallel, respecting local ownership. For the solver itself, we employed the Flexible GMRES (FGMRES) Krylov method, preconditioned by an additive Schwarz method (ASM) with overlapping subdomains. Each ASM block used a direct LU factorization to approximate the local inverse, with symbolic factorization reused to improve performance across iterations.

This hybrid shared-memory and distributed-memory approach enabled scalable and robust inversion of large, sparse, and non-symmetric systems —- while preserving the flexibility and simplicity of a single-threaded frontend for rapid experimentation and scripting.

Having the external solver code reside in an external module, which coupled directly into PETSc gave us the ability to quickly experiment with different matrix solving approaches.  The Lua interface that was used to define how our matrices were built gave us detailed control over what physical problem the solver was solving. Combining these two approaches gave us extreme flexibility to experiment with various strategies.

\vspace{1em}
\textbf{Matrix Reordering to Reduce Fill-in:} Another strategy we investigated involved reordering the system of equations to minimize matrix fill-in during factorization. In our original formulation, the system was structured in large blocks: the Poisson equations appeared first, followed by the electron continuity equations, hole continuity equations, and finally the trap state equations (for both electrons and holes). However, this block-wise ordering resulted in significant off-diagonal coupling, as physically adjacent quantities were often located far apart in the system matrix.

To address this, we implemented a configurable reordering scheme that allowed the equations to be grouped by mesh point rather than by equation type. In the revised ordering, each set of local variables -- electrostatic potential, electron and hole quasi-Fermi levels, and trap occupations -- were placed contiguously for each mesh point. This brought strongly coupled variables closer together in the matrix, reducing bandwidth and potentially lowering fill-in during LU factorization.

While this approach is promising in principle, especially for sparse direct solvers, the benefits diminished in higher dimensions. In 2D and 3D, the physical coupling between neighboring nodes naturally introduces long-range dependencies in the matrix structure, regardless of local reordering. As a result, we observed only limited improvements in solver performance. PETSc’s iterative solvers showed marginal gains in convergence rate with reordered matrices, but the overall impact was modest.

\vspace{1em}
\textbf{Matrix Normalization:} We also explored various normalization schemes to improve numerical stability and solver convergence. A straightforward approach -- scaling each matrix row to have a maximum absolute value of one -- was initially attempted. However, this method often introduced instability, particularly in cases where rows contained near-zero or zero entries. The instability likely stemmed from inconsistent scaling across physically disparate equations, as well as potential amplification of numerical noise.

As a more robust alternative, we adopted a block-wise normalization strategy based on the underlying physical equations. Specifically, we scaled the Poisson equation, electron continuity equation, hole continuity equation, and trap-state equations (for both electrons and holes) by characteristic physical constants relevant to each equation. For example, Poisson’s equation was scaled by the dielectric permittivity, while continuity equations were scaled by typical mobility or carrier density values. Although the final matrix structure no longer retained clean block separation due to our rearranged matrix, this form of normalization still maintained physical consistency across the system.

Overall, this physically motivated normalization approach provided modest improvements in solver stability and convergence. While the gains were not dramatic, this method was retained in our final implementation.

\vspace{1em}
\textbf{Final Methodology:} After extensive experimentation, we settled on a hybrid strategy that balanced physical fidelity with computational tractability. We retained the block-wise normalization scheme, as it provided physically consistent scaling across different equation types and modestly improved numerical stability. We also preserved the reordered matrix layout, placing all variables associated with a given mesh point—potential, carrier densities, and trap occupations—adjacent to one another in memory. While the performance gains were modest, this layout offered slight improvements when using PETSc-based solvers.

To enable large-scale simulations of BHJ morphologies on desktop-class hardware, we adopted a pragmatic decomposition strategy. Specifically, we performed simulations using uncoupled two-dimensional slices along one axis (e.g., fixed \( z \), sweeping over \( x\text{--}y \) planes), followed by a second simulation using orthogonal slices (e.g., fixed \( x \), sweeping over \( y\text{--}z \) planes). The resulting current densities and carrier distributions from each sweep were then averaged to produce the final output. In effect, this amounted to running two partially decoupled 3D simulations and combining the results.

To support this approach, the Newton solver was extended to manage multiple independent solver states. Once one directional sweep was completed, the solver would switch to the alternate state and repeat the process. All configuration, including slice ordering, solver control, and averaging logic, was implemented through Lua scripting. 

This strategy represented a deliberate compromise between accuracy and runtime. Our guiding principle was that simulations should complete on a standard workstation within minutes -- not hours -- to preserve an efficient experimentation and learning cycle. This constraint ruled out large-scale fully coupled 3D solves using supercomputing resources, and motivated the lightweight, flexible methodology adopted in this work.

\vspace{1em}
\textbf{Tightly Coupled Implementation:} The entire simulation framework was implemented in C, using native libraries for both Linux and Windows. This low-level, tightly coupled architecture allowed for efficient matrix construction, direct memory access, and minimal overhead in data transmission. All core components -- including Jacobian assembly, residual evaluation, and communication with solver backends -- were handled directly via API calls and memory-mapped buffers, enabling high performance and fine-grained control.

Excluding the time spent assembling the Jacobian and solving the linear system, one of the most significant computational bottlenecks was evaluating the exponential function. Exponentials are inherently expensive to compute on modern CPUs due to their reliance on transcendental instructions, which typically incur high latency.

To mitigate this, we employed several optimizations. First, the Bernoulli function was carefully implemented to reuse shared terms wherever possible, minimizing redundant evaluations. Second, we introduced a caching mechanism to store recently computed exponential values. Because neighboring mesh points often require the same or similar exponential terms—especially when sweeping across the domain—it is common for identical values to be recomputed multiple times. By caching these values in memory, we reduced redundant evaluations and improved overall performance.

These optimizations, combined with the use of C for all core routines, contributed to a highly efficient implementation capable of executing large-scale simulations on standard desktop hardware.

\vspace{1em}
\textbf{Visualization:} In addition to developing the solver itself, significant effort was devoted to visualization, which proved essential both for interpreting results and for debugging during solver development. Real-time insight into intermediate states -- such as potential distributions, carrier densities, or current flow -- can be invaluable when assessing convergence behavior or diagnosing unexpected physical effects.

To support this, we implemented a custom OpenGL-based visualization tool. The rendering engine was written in C for performance, but the user interface was built using \texttt{PySide} (Qt for Python), with communication between the two layers handled via a lightweight pipe. This architecture allowed for efficient rendering while maintaining UI flexibility and interactivity.

The viewer included dynamic controls -- such as sliders to traverse slices along each spatial axis -- enabling users to inspect 2D cross-sections of 3D fields in real time. This proved particularly useful when tuning convergence parameters, validating boundary conditions, or comparing directional solver sweeps. The combination of hardware-accelerated rendering and Python-based UI scripting offered an effective and responsive visualization workflow.

\vspace{1em}
\textbf{Benchmarking:} To assess performance and practical limits, we benchmarked the solver across a range of system sizes and configurations. For a full 3D system with a \(10 \times 10 \times 10\) mesh and four electron and four hole trap levels per mesh point (i.e., a total of eight trap states per node), the solver completed a full device simulation in under two minutes on a standard desktop workstation. Convergence was robust and independent of morphology, with no numerical instability observed.

However, as the mesh size increased, the problem quickly became intractable using fully coupled Newton--Raphson iterations. At \(20 \times 20 \times 20\), convergence became significantly more challenging and solver runtimes increased sharply. Fully coupled simulations for systems larger than \(30 \times 30 \times 30\) were generally impractical due to excessive memory use and convergence failures. Simulations at \(50^3\) were effectively impossible using the monolithic solver.

To overcome these limitations, we relied on directional slicing strategies, which are described above. A typical simulation using the two-pass averaging scheme -- i.e., uncoupled 2D slice sweeps along orthogonal directions followed by current averaging -- allowed for \(30 \times 30\) simulations with realistic BHJ morphology. A complete current--voltage (JV) sweep from 0 V to 1.1 V, in voltage steps of 0.02 V, could be completed in approximately 30 minutes using this approach. This represented a practical compromise between resolution, accuracy, and runtime suitable for iterative device design and morphological exploration.

\vspace{1em}
\textbf{Unexplored Strategies:} Due to time constraints and prioritization of robustness, several potentially useful strategies were not explored in this work. One such idea involves spatially selective equation reduction -- specifically, avoiding the solution of drift--diffusion equations for minority carriers in regions where their densities are negligible.

In our current implementation, drift--diffusion equations are solved uniformly throughout the entire device, including deep within donor- or acceptor-rich domains where one carrier type dominates. While this approach is physically well-motivated -- since both carrier types influence the electrostatic potential everywhere -- it may be computationally inefficient. Omitting the minority carrier equations in regions of low relevance could, in principle, reduce the number of unknowns by up to 50\%, leading to smaller matrices and faster solve times.

However, implementing such a scheme would significantly complicate the matrix structure. Selectively removing equations introduces sparsity asymmetry and irregular coupling patterns, which could increase matrix fill-in and destabilize the solver. Electrostatic consistency may also suffer if boundary conditions between full and reduced regions are not handled carefully. 

Given these trade-offs, and based on prior experience where substantial implementation effort did not always yield performance improvements, we chose not to pursue this path. Nonetheless, spatially adaptive equation reduction remains a potential area for future exploration, particularly in large-scale or morphology-aware simulations.

\subsection{Electrical paramters}
\begin{table}[ht]
\centering
\begin{tabular}{llr}
\hline
\textbf{Parameter} & \textbf{Symbol} & \textbf{Value (Unit)} \\
\hline
Electron mobility (in acceptor) & $\mu_n$ & $1 \times 10^{-5}~\mathrm{m}^2\mathrm{V}^{-1}\mathrm{s}^{-1}$ \\
Hole mobility (in acceptor) & $\mu_p$ & $1 \times 10^{-10}~\mathrm{m}^2\mathrm{V}^{-1}\mathrm{s}^{-1}$ \\
Electron mobility (in donor) & $\mu_n$ & $1 \times 10^{-10}~\mathrm{m}^2\mathrm{V}^{-1}\mathrm{s}^{-1}$ \\
Hole mobility (in donor) & $\mu_p$ & $1 \times 10^{-5}~\mathrm{m}^2\mathrm{V}^{-1}\mathrm{s}^{-1}$ \\
Effective density of states (electrons) & $N_{c}$ & $5 \times 10^{25}~\mathrm{m}^{-3}$ \\
Effective density of states (holes) & $N_{v}$ & $5 \times 10^{25}~\mathrm{m}^{-3}$ \\
Electron trap density & $N_\mathrm{t,e}$ & $1 \times 10^{24}~\mathrm{m}^{-3}\mathrm{eV}^{-1}$ \\
Hole trap density & $N_\mathrm{t,h}$ & $1 \times 10^{24}~\mathrm{m}^{-3}\mathrm{eV}^{-1}$ \\
Electron tail slope & $E_\mathrm{tail,e}$ & $50~\mathrm{meV}$ \\
Hole tail slope & $E_\mathrm{tail,h}$ & $50~\mathrm{meV}$ \\
Free $\rightarrow$ trapped x-section (e$^-$) & $\sigma_{n}^{\mathrm{trap}}$ & $1 \times 10^{-20}~\mathrm{m}^2$ \\
Trapped $\rightarrow$ free x-section (e$^-$) & $\sigma_{n}^{\mathrm{detrap}}$ & $1 \times 10^{-21}~\mathrm{m}^2$ \\
Free $\rightarrow$ trapped x-section (h$^+$) & $\sigma_{p}^{\mathrm{trap}}$ & $1 \times 10^{-21}~\mathrm{m}^2$ \\
Trapped $\rightarrow$ free x-section (h$^+$) & $\sigma_{p}^{\mathrm{detrap}}$ & $1 \times 10^{-20}~\mathrm{m}^2$ \\

Electron affinity & $\chi$ & $1.6~\mathrm{eV}$ \\
Bandgap & $E_\mathrm{g}$ & $1.2~\mathrm{eV}$ \\
Scattering length & $L_\mathrm{scatt}$ & $1 \times 10^{-8}~\mathrm{m}$ \\
Lifetime & $\tau$ & $5 \times 10^{-10}~\mathrm{s}$ \\
Radiative decay rate & $k_r$ & $4 \times 10^{9}~\mathrm{s}^{-1}$ \\
Dissociation rate constant & $k_\mathrm{diss}$ & $1 \times 10^{11}~\mathrm{s}^{-1}$ \\
Bimolecular rate constant & $k_\mathrm{alpha}$ & $1 \times 10^{-13}~\mathrm{m}^3\mathrm{s}^{-1}$ \\
\hline
\end{tabular}
\caption{Simulation parameters}
\label{tab:sim_params}
\end{table}

Chemical diagrams in Figure 1 drawn after \cite{Woepke2022transport}.
\end{document}